\documentclass[bibyear]{corot}  

\usepackage{graphicx}
\usepackage{txfonts}
\usepackage{rotating}

\begin{document}

   \title{Exploration of the brown dwarf regime around solar-like stars by CoRoT}


   \author{Sz. Csizmadia
          \inst{1}
          }

   \institute{DLR, Institut f\"ur Planetenforschung, 12489 Berlin, Rutherfordstr 2, Germany, \\
              \email{szilard.csizmadia@dlr.de}
             }

 
  \abstract
   {}
   {A summary of the CoRoT brown dwarf investigations are presented.}
   {Transiting brown dwarfs around solar like stars were studied by using the
   photometric time-series of CoRoT, and ground based radial velocity measurements.}
   {CoRoT detected three transiting brown dwarfs around F and G dwarf stars. The 
   occurence rate of brown dwarfs was found to be $0.20\pm0.15$\% around solar-like
   stars which is compatible with the value obtained by Kepler-data.}
   {}


   \maketitle
%

\section{Introduction}

How shall we classify brown dwarfs? Are they gaseous giant planets as Hatzes \& Rauer (2015) 
suggest, or are they stars as we can find in Trentham et al (2001), or do they 
form a separate class of celestial objects in their own right?

Beyond the classification problem, their role is not well understood in planet 
formation, in evolution of planetary systems, in the chemical and dynamical 
evolution of the Galaxy; their impact on habitability is not well known both as 
host objects as well as additional objects in a planetary system; they can have 
moons whose habitability is not clear; and they are not studied well enough as 
planet hosts, although they can harbour planets up to about 5 Earth-masses 
(Payne \& Lodato 2007).

When the minimum mass-limit to hold nuclear fusion was investigated some decades 
ago, it was found that a hydrogen gas sphere over 80 $M_{\mathrm{jup}}$ has enough mass 
to hold nuclear fusion -- fusion of hydrogen, helium or heavier elements -- for 
millions to billions of years, so they are called stars; below 13 $M_{\mathrm{jup}}$ we 
find the regime of planetary objects who exhibit no natural fusion at all. Between 
these limits sit brown dwarfs who fuse deuterium (D) or lithium (Li) to helium-3 
and helium-4, respectively, but only for typically $\sim0.1$~Myrs which sounds a very 
episodic event in their life-cycle because after this phase they simply cool down 
and contract (Baraffe et al. 2003). Their evolution after this event is quite 
similar to the contraction of Jupiter-like gas planets. This is not surprising 
because they consist of mostly hydrogen-forming degenarate electron gas core 
inside and this structure, of course, resembles the structure of gas giant 
planets. The modern lower and upper limits for brown dwarfs are 11--16 $M_{\mathrm{jup}}$ 
depending in exact internal chemical composition (Spiegel et al. 2011) and 75--80 
$M_{\mathrm{jup}}$ (Baraffe et al. 2002), respectively.

The number of known brown dwarfs is over 2200 and more than 400 are in binary 
systems, the rest are single (Johnston 2015). Only 66 brown dwarfs orbit their host 
stars on closer orbit than 2 au (65 are listed in Ma \& Ge 2014 and we extended  their
list with CoRoT-33b, see Csizmadia et al. 2015). Only 12 or 13 transiting bona fide brown
dwarfs are known (Table 1), depending on how we count the status of  KOI-189b whose
mass is 78$\pm$5 Jupiter-masses, so it is at the brown dwarf - star
boundary. There are at least two eclipsing binary systems known in  which two brown
dwarfs orbit each other (see Table 1). At least one eclipsing  binary system is known
where a brown dwarf orbits an M-dwarf (NLTT 41135Ab,  Irwin et al. 2010). The
remaining eclipsing brown dwarfs orbit solar-like stars  (defined as main sequence or
just slightly evolved FGK stars). For the study of their evolution, internal
structure and impacts in the aforementioned issues we need very precise and
model-independent mass, radius, luminosity and chemical composition values. 
Transiting systems can provide mass and radius with good precision and therefore 
they are important cases.

Here we summarize the contribution of CoRoT to the development of our knowledge of 
brown dwarfs, especially of those which are in binary systems around solar-like
stars.

%

\section{Brown dwarfs found by CoRoT}

CoRoT has reported three brown dwarf discoveries. This is comparable to the four
brown dwarfs detected by Kepler (see Table 1) since both mission observed about the
same number of stars (ca. 150,000) but for different time-intervals (30--150 days vs
four years). Notice that the detection bias for Jupiter-sized objects are the same at
short periods and in this size-range. Therefore it is not surprising that the derived
brown dwarf occurence rates were found similar by both space missions (Csizmadia et
al. 2015; Santerne et al. 2015). Hereafter we discuss the individual cases one by one
and in the next section we concentrate on the occurence rates.

\subsection{CoRoT-3b: the first habitant of the brown dwarf desert}

Brown dwarfs, as companion objects in binary systems, exhibit a much smaller
occurence rate than stars and planets and this is called the 'brown dwarf desert'. It was
found and confirmed by the radial velocity method (Marcy \& Butler 2000; Lafreni\`ere et
al. 2007; Patel et al. 2007; Wittenmyer et al. 2009; Sahlmann et al. 2011), as well as
by adaptive optics direct imaging (Metchev \& Hillenbrand 2009).

The discovery of CoRoT-3b definitely means a breakthrough and a significant 
milestone in brown dwarf research (Deleuil et al. 2008). It was the first  object
in the brown dwarf desert whose mass and radius were measured from its 
transiting nature. It was also a surprise because nobody expected the existence of a
brown dwarf so close -- only 7.8 stellar-radii -- to a solar-like star ($P_{orb} =
4.26$ days) because radial velocity surveys did not detect any similar object before
the discovery of CoRoT-3b (only two suspected objects between a minimum mass of 10
and 20 Jupiter-masses were known at that time, see Deleuil et al. 2008). Therefore it
was not clear then, that CoRoT-3b is a brown dwarf or a 'super-planet'. A
super-planet could be formed via core-accretion whose mass can be up to 25 $M_{\mathrm{jup}}$ 
without deuterium-burning or such a high-mass object can be formed via collision of
several smaller planetesimals or planets (Deleuil et al. 2008 and references
therein). The origin of CoRoT-3b is still under debate. Notice that model
calculations of Mordasini et al. (2009) predict that high mass planets and brown
dwarfs can form up to 40 $M_{\mathrm{jup}}$ via core-accretion but none of these objects get
closer than 1 au to their host star (cf. Fig. 9 of Mordasini et al. 2009). Although
Armitage \& Bonnell (2002) proposed a very effective migration process for brown
dwarfs, it is questionable that it is really so effective that the majority of them are
engulfed by their host stars and this is the cause of the rarity of close-in
brown dwarfs. Therefore, if CoRoT-3b ($\sim22~M_{\mathrm{jup}}$) and NLTT 41135b ($\sim34
M_{\mathrm{jup}}$) formed via core-accretion, then it is an intriguing question how they moved 
so close to their host stars.

At the time of the detection of CoRoT-3b, the authors thought that this object
confirms the suspicion that ''transiting giant planets ($M > 4 M_{\mathrm{jup}}$) can be found
preferentially around more massive stars than the Sun'' (Deleuil et al 2008). The
discovery of NLTT 41135b, a $\sim34 M_{\mathrm{jup}}$ gaseous giant planet (or a brown dwarf)
around an M-dwarf is a remarkable counter-example (Irwin et al. 2010).

\subsection{CoRoT-15b, an oversized brown dwarf}

CoRoT-15b was the second detected transiting brown dwarf (Bouchy et al. 2011a). Its most exciting feature 
is its high radius relative to its mass ($1.12^{+0.30}_{-0.15}~R_{\mathrm{jup}}$, a mass of 
$63.3\pm4.1~M_{\mathrm{jup}}$). Notice that brown dwarfs contract slowly until the 
equilibrium size in most of their lifetime: but at the beginning they can have as 
large radius as 4-5 Jupiter radii, but at the age of the Universe and at the mass 
of CoRoT-15b the radius should be around 0.8 Jupiter-radii. The radius-variation is 
very fast in the first 5 Gyrs (Baraffe et al. 2003). The estimated age of the host 
star is between 1.14--3.35 Gyr using STAREVOL and $1.9\pm1.7$ Gyr obtained by CESAM. 
These do not contradict each other but the age is not well constrained - this 
is a point where PLATO with its well-measured (better than 10\%) ages will play a 
role in the study of brown dwarfs and of planets (Rauer et al. 2014). 
Therefore Bouchy et al. (2011a) left the question open as to why this brown dwarf has 
large radius: either the system is young, or cold spots on the brown dwarf surface 
help to inflate the radius or atmospheric processes blow up it with disequilibrium 
chemistry. The irradiation effects were found to be negligible in the 
inflation-process of brown dwarfs.

Interestingly, CoRoT-15b may be a double-synchronous system: the orbital period of 
the brown dwarf and the rotational period of the star can be equal to each other, 
but the precision of the rotational period measurement is not enough to make this 
statement conclusive (Bouchy et al. 2011a). If subsequent investigation can confirm this 
suspicion, then it seems that tidal interaction between stars and close-in brown 
dwarfs are strong enough to synchronize their stars or trap it in some resonance. 
We further discuss this in the light of CoRoT-33b.

\subsection{CoRoT-33b, a key object for tidal evolution}

This system was reported in Csizmadia et al. (2015). 
CoRoT-33A could be the presently-known most metal-rich brown dwarf host star 
because its metallicity is a $[Fe/H] = +0.44\pm0.1$ (the other candidate 
for this title is HAT-P-13A with $[Fe/H] = +0.43\pm0.1$, see Bakos et al. 2009). 
The host stars of the brown dwarfs in binary systems seem to be metal-poor 
(Ma \& Ge 2014), therefore this system helps to extend the sample to the tails of 
the distribution.

The host star seems to be older than 4.6 Gyr and likely it is even older (maybe as  old as 11-12 Gyr). The rotation period
is too small for a G9V star when we compare it to the braking-mechanisms of the single-star scenarios of Bouvier et al.
(1997). The measured $v \sin i$, stellar radius, as well as the observable spot modulation on the light curve show that
$P_{rot} = 8.95$~days.  Another interesting feature of CoRoT-33b, whose mass is 59 $M_{\mathrm{jup}}$, is its eccentric
orbit ($e = 0.07$). Since the orbital period is just 5.82 d, the circularization time-scale for such a system is much
shorter than the age of the system. Even more interestingly, the orbital period of the brown dwarf is within 3\% of a 3:2
commensurability with the rotational period of the star. 

B\'eky et al. (2014) listed six hot Jupiters where strange commensurabilities 
can be observed between the stellar rotational rate and orbital periods of the 
planetary companions. They also suspected that these are just random coincidences 
between the stellar rotational period and the planetary companions' orbital periods 
(maybe a stellar spot at a certain latitude may mimic such a coincidence due to the 
differential rotation of the star). However, several host stars of brown dwarfs 
rotate faster -- even if we take into account the normal rate for stellar 
differential rotation -- than we expect from their ages, so such random coincidences 
cannot explain the observed phenomena in star-brown dwarf systems in general. More 
likely, we see a long-term interaction between the star and the close-in, 
high mass brown dwarf system. This interaction consists of tidal interaction as 
well as magnetic braking effects. Such a combination of star - planet/brown dwarf 
interaction can explain the observed properties -- even the eccentric orbit -- of 
CoRoT-33 and other systems. Details of this physical mechanism and the results can 
be found in Ferraz-Mello et al. (2015).

\section{Distance-occurance rate relationship?}

The CoRoT data allowed us to determine the relative frequency ratio of brown dwarfs to hot Jupiters in the $P<10$ days
orbital period range. Using the true frequency of hot Jupiters as given in Wright et al. (2012), an $0.20\pm0.15$\% true
occurence rate of brown dwarfs was found around solar-like stars for $P<10$ days (Csizmadia et al. 2015). It is also
suspected that this occurence rate follows a power-law up to at least 1000 au orbital separations: 
\begin{equation} 
f = \alpha \left( \frac{a}{1\mathrm{au}} \right)^{\beta} 
\end{equation} 
where $f$ is the occurence rate of brown dwarfs around
solar like stars below ~1000  au orbital separation and first estimates give, $\alpha=0.55^{+0.8}_{-0.55}$\%,  $\beta =
0.23 \pm 0.06$ (Csizmadia et al 2015). The occurence rate-separation  relationship considers radial velocity, microlensing
and direct imaging  results,  too (see the discussion and references in Csizmadia et al. 2015). Although  ground-based
transit surveys like HAT, WASP etc. did not report this frequency rate  so far, the analysis of Kepler data is fully
compatible with the CoRoT-results and  also supports the aformentioned relationship (Santerne et al. 2015). The meaning of 
this possible relationship and its connection to formation theories of planetary  systems is not studied yet. However, it
is worth mentioning that models by  Mordasini et al. (2009) predicted the formation of brown dwarfs via core accretion  up
to 40 $M_{\mathrm{jup}}$ but none of these objects get closer than 1 au according to  their Fig. 9. Observational results
of  CoRoT, Kepler and ground based surveys  (Table 1) shows that somehow these brown dwarfs moved inward significantly.

\section{Summary and future prospects for transiting brown dwarf hunting}

CoRoT detected three transiting brown dwarfs, including the first known such object.
All three are very close to their host stars ($P_{orb}<10$~days). Two of them
(CoRoT-15b and -33b) show interesting commensurabilities between the orbital period
of the transiting object and the rotational period of the host star (maybe 1:1 in the
case of 15b and a strict 3:2 in the case of 33b). Well-measured masses and first
estimates of the radii were reported. CoRoT-33b also has an eccentric orbit and all
three objects can be subject of future tidal evolution studies. The occurence rate of
brown dwarfs was estimated for the ten days orbital period range and it was found to
be 0.2\%$\pm$0.15\% and this was confirmed by an analysis of the Kepler-data later
(Santerne et al. 2015). The presence of such close-in brown dwarfs is a challange for
presently known formation theories.

Transiting brown dwarfs are gold-mines for their studies. The mass and radius 
(hence their mean density) can be measured in a model-independent way for them, and 
the random and systematic uncertainties of their parameters in such binary systems 
are dominated mostly by the stellar parameters (in double-lined systems this kind 
of uncertainty does not appear). This will be improved by the next generation of 
instruments which will be more sensitive for secondary eclipses and phase curves, 
like PLATO. Since the age can be measured by isochrone fitting 
or by asteroseismology in the future (Rauer et al. 2014), the predicted contraction 
rate of brown dwarfs and thus the theoretical models of them (Baraffe et al. 2003) 
can be checked.

Although almost a dozen transiting brown dwarfs are known, this is still too small 
a sample for such studies. Since the size of brown dwarfs is in the Jupiter-sized 
range or it can be bigger (up to several Jupiter-radii) for young ones, they can 
easily be detected from ground, too. However, interestingly, several brown dwarfs 
are grazing transiters (like NLTT 41135b or CoRoT-33b) and that decreases the 
observed transit depth making the discovery hard from ground. For some yet unkown
reason, space observatories detected higher brown dwarf/hot Jupiter ratio than what
we can suspect from ground based surveys if we simply divide the number of the
observed brown dwarfs by the number of hot Jupiters. It is quite unlikely that space
observatories missed hot Jupiters, more probably this may be a selection effect of
the ground based surveys.

We foresee several ongoing or planned space missions which are able to detect  
transiting brown dwarfs, like Gaia (launched 2013), CHEOPS and TESS (to be launched 
2017), PLATO (to be launched in 2024). Also, EUCLID (to be launched in 2022) may 
detect a limited number of microlensing brown dwarfs as a by-product.  Gaia is also
able to detect brown dwarfs via its primary technique, namely via astrometry.

CHEOPS targets known planets and candidates detected by radial velocity  
technique (RV). CHEOPS may search for the possible transits of these RV-detected 
objects which would allow to determine the inclination and hence their true masses 
instead of a lower mass limit; also, their radius becomes known. One can propose 
that CHEOPS may extend its program by checking the possible transits of RV-detected 
brown dwarf candidates.

There also are several ongoing ground-based surveys which are able to find 
transiting (NGTS, WASP, HAT, for instance) or microlensing brown dwarfs. However, 
the low efficiency or observational biases of ground based survey are hard to 
understand and requires further study and a careful check of the existing data 
for undetected brown dwarfs. The same is to apply to space-based observatories' 
data.

%

\begin{sidewaystable*}
\caption{Basic data of known transiting brown dwarfs. $\rho$ is the mean density of 
the brown dwarf component. Below the line one can find the questionable systems. 
Periods are truncated after the third decimal place. This table is an extended and updated
version of the one published in Csizmadia et al. (2015).}
\centering
\begin{tabular}{l@{\hskip1mm}l@{\hskip1mm}c@{\hskip1mm}c@{\hskip1mm}c@{\hskip1mm}c@{\hskip1mm}c@{\hskip1mm}c@{\hskip1mm}c@{\hskip1mm}c@{\hskip1mm}c@{\hskip1mm}c@{\hskip1mm}c@{\hskip1mm}}
\hline\hline                 
Name  & Mag & $M_\mathrm{star}/M_\odot$ & $R_\mathrm{star}/R_\odot$ & $T_\mathrm{star}$ [K] & $\mathrm{[Fe/H]}$ & $P$ (days) & $e$ & $M_\mathrm{BD}/M_\mathrm{Jup}$ &$R_\mathrm{BD}/R_\mathrm{Jup}$ & $\rho$ [g/cm$^3$] & Ref. \\
\hline                                         			  
2M0535-05a$^a$&  19.21R&                           &                        &                  &                   & $9.779  $ & 0.3225$\pm$0.0060           & $56.7\pm4.8$            & 6.5$\pm$0.33              & 0.26$\pm$0.06     & 1 \\
2M0535-05b$^a$&        &                           &                        &                  &                   & $9.779  $ & 0.3225$\pm$0.0060           & $35.6\pm2.8$            & 5.0$\pm$0.25              & 0.35$\pm$0.08     & 1 \\
CoRoT-3b      &  13.29V& $1.37\pm0.09$             & $1.56\pm0.09$          & $6740\pm140$     & -0.02$\pm$0.06$^b$& $4.256    $ & 0.0                         & $21.66\pm1.0$           & $1.01\pm0.07$             & 26.4$\pm$5.6      & 2 \\
CoRoT-15b     &  15.4R & $1.32\pm0.12$             & $1.46^{+0.31}_{-0.14}$ & $6350\pm200$     & +0.1$\pm$0.2      & $3.060    $ & 0                           & $63.3\pm4.1$            & $1.12^{+0.30}_{-0.15}$    & 59$\pm$29         & 3 \\
CoRoT-33b     &  14.25R& $0.86\pm0.04$             & $0.94^{0.14}_{-0.08}$  & $5225\pm80$      & +0.44$\pm$0.10    & $5.819$   & $0.0700\pm0.0016$ 	     & $59.0^{+1.8}_{-1.7}$    & $1.10\pm0.53$             & 55$\pm27$         & 4 \\
KELT-1b       &  10.63V& $1.335\pm0.063$           & $1.471^{+0.045}_{-0.035}$& $6516\pm49$    & +0.052$\pm$0.079  & $1.217$   & $0.01^{+0.01}_{-0.007}$     & $27.38\pm0.93$          & $1.116^{+0.038}_{-0.029}$ & $24.5^{1.5}_{-2.1}$ & 5 \\
Kepler-39b$^c$&  14.43R& $1.10^{+0.07}_{-0.06}$    & $1.39^{+0.11}_{-0.10}$ & $6260\pm140$     & -0.29$\pm$0.10    & $21.087$     & $0.121^{+0.022}_{-0.023}$   & $18.00^{+0.93}_{-0.91}$ & $1.22^{+0.12}_{-0.10}$    &12.40$^{+3.2}_{-2.6}$ & 6 \\
Kepler-39b$^c$&        & $1.29^{+0.06}_{-0.07}$    & 1.40$\pm$0.10        & $6350\pm100$       & +0.10$\pm$0.14    & $21.087$  & $0.112\pm0.057$             & $20.1^{+1.3}_{-1.2}$    & $1.24^{+0.09}_{-0.10}$    &13.0$^{+3.0}_{-2.2}$ & 7 \\
KOI-189b$^d$  &  12.29K& $0.764\pm0.051$           & 0.733$\pm$0.017        & $4952\pm40$ 	   & -0.07$\pm$0.12 & $30.360$  & $0.2746\pm0.0037$           & $78.0\pm3.4$            & $0.998\pm0.023$           & 97.3$\pm$4.1        & 8 \\
KOI-205b      &  14.47i& $0.925\pm0.033$	   & $0.841\pm0.020$        & $5237\pm60$      & +0.14$\pm$0.12    & $11.720$ & $<$0.031                    & $39.9\pm1.0$            & $0.807\pm0.022$  	   & 75.6$\pm$5.2        & 9 \\
KOI-205b      &        & $0.96^{+0.03}_{-0.04}$    & $0.87\pm0.020$         & $5400\pm75$      & +0.18$\pm$0.12    & $11.720$  & $<$0.015                    & $40.8^{=1.1}_{-1.5}$    & $0.82\pm0.02$  	   & 90.9$^{+7.2}_{6.8}$ & 6 \\
KOI-415b      &  12.66K& $0.94\pm0.06$             & $1.15^{+0.15}_{-0.10}$ & $5810\pm80$      & -0.24$\pm$0.11    & $166.788$  & $0.698\pm0.002$             & $62.14\pm2.69$          & $0.79^{+0.12}_{-0.07}$    &157.4$^{+51.4}_{-52.3}$& 10 \\
LHS 6343C$^e$ &  13.88V&$0.370\pm0.009$            & $0.378\pm0.008$        & $3130\pm20$      & +0.04$\pm$0.08    & $12.713$    & $0.056\pm0.032$             & $62.7\pm2.4$            & $0.833\pm0.021$           &109$\pm$8            & 11 \\
LHS 6343C$^e$ &        &$0.381\pm0.019$            & $0.380\pm0.007$        & $3431\pm21$      & +0.03$\pm$0.26    & $12.713$  & $0.030\pm0.002$             & $64.6\pm2.1$            & $0.798\pm0.014$           &170$\pm$5            & 17 \\
WASP-30b      &  11.91V& $1.166\pm0.026$           & $1.295\pm0.019$        & $6201\pm97$      & -0.08$\pm$0.10    & $4.156$   & 0                           & $60.96\pm0.89$          & $0.889\pm0.021$           &107.6$\pm$1.1        & 12 \\
NLTT 41135b   &   8.44V& $0.188^{0.026}_{0.022}$   & $0.21\pm0.015$         & $3230\pm130$     & 0.0               & $2.889$   & n/a                         & $33.7\pm2.8$            & $1.13_{-0.17}^{+0.27}$    &$29^{+23}_{-15}$     & 14 \\
EPIC 2038a$^i$&  &                           &                        &                  &                   & $4.451$   &$0.3227\pm0.0042$  & $23.22\pm0.47$ & $2.75\pm0.05$ & $1.38\pm0.08$ & 18 \\
EPIC 2038b$^i$&  &                           &                        &                  &                   & $4.451$   &$0.3227\pm0.0042$  & $25.79\pm0.58$ & $2.485\pm0.004$ & $2.09\pm0.07$ & 18 \\
\hline
1SWASP J1407b$^f$&12.4V& 0.9                       & $0.99\pm0.11$          & $4400^{+100}_{-200}$&  n/a           & $3725\pm900$     & n/a                         & $20\pm6$                & n/a                       &n/a                  & 13 \\ 
Kepler-27c$^g$&     n/a& $0.65\pm0.16$             & $0.59\pm0.15$          & $5400\pm60$      & $0.41\pm0.04$     & 31.330          & n/a                         & $<13.8^h$               & 0.44                   &n/a                  & 15 \\
Kepler-53b$^g$&   16.0V& 0.89                      & 0.98                   & 5858             & n/a               & 18.648 &n/a                          & $<18.41^h$              & 0.26                      &n/a                  & 16 \\
Kepler-53c$^g$&   16.0V& 0.89                      & 0.98                   & 5858             & n/a               & 38.558 &n/a                          & $<15.74^h$              & 0.28                      &n/a                  & 16 \\
Kepler-57b$^g$&   15.5V& 0.83                      & 0.73                   & 5145             & n/a               & 5.729  &n/a                          & $<18.86^h$              & 0.195                     &n/a                  & 16 \\
\hline
\end{tabular}
\tablefoot{References: 
1: Stassun et al. (2006), 
2: Deleuil et~al. (2008),  
3: Bouchy et~al. (2011a),  
4: Csizmadia et al. (2015) 
5: Siverd et~al. (2012),   
6: Bouchy et~al. (2011b),  
7: Bonomo et al. (2015),   
8: D\'{\i}az et~al. (2014), 
9: D\'{\i}az et~al. (2013), 
10: Moutou et~al. (2013),   
11: Johnson et~al. (2011),  
12: Anderson et~al. (2011),  
13: Kenworthy et al. (2014), 
14: Irwin et al. (2010),     
15: Steffen et al. (2012),   
16: Steffen et al. (2013),   
17: Montet et al. (2015),    
18: David et al. (2015)     
}
\tablefoot{
{\it a:} 2M0535-05 is an extreme young eclipsing system in which two brown dwarfs orbit each other. Identical to V2384 Orionis.
{\it b:} $\mathrm{[M/H]}$ value is reported in the reference. Notice that $[M/H] \approxeq [Fe/H]$; we did not convert the inhomogeneous [Fe/H] to the same scale.
{\it c:} aka KOI-423b.
{\it d:} D\'{\i}az et al. (2014) concluded that KOI-189b can be either a high-mass brown dwarf or a very low mass star, too, therefore its status is uncertain.
{\it e:} the brown dwarf orbits companion A of a binary system, 
and data of the component A is given here. Star B has $M=0.30\pm0.01\mathrm{M}_\odot$, 
$T_\mathrm{eff}=3030\pm30$~K (Johnson et~al. 2011).
{\it f:} the host star is young (16 Myr) and surrounded by a multiring-structured (protoplanetary?) disc, see Mamajek et al. (2012).
{\it g:} masses are TTV masses, not RV. 
{\it h:} upper limit.
{\it i:} full name is EPIC 203 868 608. Similar system to 2M0535-05: two brown dwarfs orbit each other.
}
\label{knownBrownDwarfs}   
\end{sidewaystable*}


\end{document}